\documentclass{elsart}

    \usepackage{latexsym,bm,amsmath,amssymb,amsfonts}
    \usepackage{epsfig,graphics,graphicx}

\long\def\comment#1{ }

\newcommand{\beq}{\begin{eqnarray}}
\newcommand{\eeq}{\end{eqnarray}}
\newcommand{\be}{\vspace{-.4cm}\begin{eqnarray}}
\newcommand{\ee}{\vspace{-.5cm}\end{eqnarray}}

\newcommand{\eq}{Eq.~}

\newcommand{\Tr}{{\rm Tr}}

\def\simge{\mathrel{%
   \rlap{\raise 0.511ex \hbox{$>$}}{\lower 0.511ex \hbox{$\sim$}}}}
\def\simle{\mathrel{
   \rlap{\raise 0.511ex \hbox{$<$}}{\lower 0.511ex \hbox{$\sim$}}}}

\newcommand{\nr}[1]{(\ref{#1})}
\newcommand{\lqcd}{\Lambda_{\mathrm{QCD}}}

\newcommand{\qs}{Q_{\mathrm{s}}}
\newcommand{\ra}{R_A}

\newcommand{\as}{\alpha_{\mathrm{s}}}

\newcommand{\dtau}{{\partial_{\tau}}}

\newcommand{\ud}{\mathrm{d}}
\newcommand{\xt}{\mathbf{x}_T}

\newcommand{\kt}{{\mathbf{k}_T}}

\begin{document}

\begin{flushright}
~\vspace{-1.25cm}\\
{\small\sf BNL--NT--06/10}\\
{\small\sf hep-ph/0602189}
\end{flushright}
\vspace{2.cm}

\begin{frontmatter}

\parbox[]{16.0cm}{ \begin{center}
\title{Some Features of the Glasma}

\author{T.~Lappi$^{\rm a}$},
\author{L.~McLerran$^{\rm a,b}$},

\address{$^{\rm a}$ Physics Department, Brookhaven
National Laboratory, Upton, NY 11973, USA}

\address{$^{\rm b}$ RIKEN BNL Research Center, Brookhaven National 
Laboratory,
Upton, NY 11973, USA}

\date{March 3rd, 2006}
\vspace{0.8cm}
\begin{abstract}

We discuss high energy hadronic collisions within the
theory of the Color Glass Condensate.  We point out that the initial electric
and magnetic fields produced in such collisions are longitudinal. This
leads to a novel string like description of the collisions, and a large
Chern-Simons charge density made immediately after the collision. The
presence of the longitudinal magnetic field suggests that essential to the
description of these collisions is the decay of Chern-Simons charge.
\end{abstract}

\end{center}}

\end{frontmatter}
\newpage

\section{Introduction}

High energy hadronic collisions can be described as collisions of sheets
of Colored Glass Condensate \cite{Kovner:1995ja,Kovner:1995ts,%
Gyulassy:1997vt,Kovchegov:1997ke,Krasnitz:1998ns,Krasnitz:1999wc,%
Krasnitz:2000gz,Krasnitz:2001qu,Lappi:2003bi}.  
The degrees of freedom of the Color Glass
Condensate are those of high energy density gluonic fields
\cite{Gribov:1984tu,Mueller:1985wy,McLerran:1994ni,McLerran:1994ka,%
McLerran:1994vd,Iancu:2000hn,Ferreiro:2001qy}. 
Because the
physical density of gluons becomes large, the typical separation between
gluons is small, and therefore $\as$ is small. The highly coherent gluons 
fill the phase space up to the maximal occupation number $\sim 1/\as$
and can thus be thought of as condensed.
Because of their high speed and Lorentz time dilation, 
the valence degrees of freedom 
are seen  by the low $x$ fields as slowly evolving in lightcone time.
Systems which evolve over long time 
scales compared to natural ones are glasses.  Hence the name Color Glass 
Condensate. It 
is possible to describe this system from first principles in QCD because 
the coupling is weak.

The equations which describe the color fields in a 
high energy hadronic collision have been
written down in Refs.~\cite{Kovner:1995ja,Kovner:1995ts,Gyulassy:1997vt,%
Kovchegov:1997ke}.  For the case where the rapidity dependence of the fields
can be ignored, the equations were numerically solved \cite{Krasnitz:1998ns,%
Krasnitz:1999wc,Krasnitz:2000gz,Krasnitz:2001qu,Lappi:2003bi}.  A successful
phenomenology has been applied to RHIC energy collisions of nuclei and dA
collisions \cite{Dumitru:2001ux,Blaizot:2004wu,Blaizot:2004wv,%
Kharzeev:2002pc,Jalilian-Marian:2003mf,Kharzeev:2003wz,Kharzeev:2004yx,%
Baier:2003hr,Albacete:2003iq,Iancu:2004bx,Dumitru:2005gt}. 
Since then, a literature has developed relating this
description to issues such as topological charge generation 
\cite{Kharzeev:1998kz,Kharzeev:1999cz,Kharzeev:2001vs,Kharzeev:2001ev}
and plasma instabilities 
\cite{Mrowczynski:1994xv,Mrowczynski:1996vh,Arnold:2003rq,%
Romatschke:2003ms,Romatschke:2004jh,Romatschke:2005pm}.

Topological charge generation is related to a number of non-perturbative
phenomena in field theory.  In electroweak theory, it leads to the anomalous
generation of baryon plus lepton number.  In QCD, it is related to
the violation of U(1) chiral symmetry, which may ultimately drive the
breakdown of the SU(2) chiral symmetry and be responsible for mass
generation in QCD.  As the topological charge is CP odd, it may also make
large CP violating fluctuations in heavy ion 
collisions \cite{Kharzeev:1998kz,Kharzeev:1999cz,Kharzeev:2001vs,%
Kharzeev:2001ev}. 
Some indications of these fluctuations have been recently observed
by the STAR expertiment~\cite{Selyuzhenkov:2005xa}.
 
The instabilities may generate the rapid thermalization seen at
RHIC.  This intermediate matter is highly coherent, and makes the
transition from the Color Glass Condensate to the Quark Gluon Plasma. We
shall call it the Glasma.  This is because the fields are coherent and 
have strength $1/g$, and so in their interaction with particle degrees of
freedom, are $\mathcal{O}(1)$, since the $g$ of interacting with the field is 
canceled  by the $1/g$ from the strength of the field.  This enhances the 
rate at which the system can rearrange in phase space and reach a 
thermalized distribution.

In this paper, we will discuss some properties of the Glasma.  Our intent 
is not to add anything to the existing mathematical literature concerning 
the solution to the classical field equations resulting from the 
collision.  In fact, we largely restrict ourselves to the somewhat tame 
situation where the fields are boost invariant, and we only briefly 
discuss plasma instabilities.

Our goal is to provide a qualitative and intuitive understanding of the
Glasma fields shortly after collisions.  This work is an elaboration and
interpretation of some of the recent observations of Fries et. al. 
\cite{Fries:2005yc,friesinprep}. The
structure we illuminate is amusing.  An infinitesimal time
after the collision, longitudinal electric and magnetic fields are
produced.  There are no transverse fields, except on the valence sheets of
charge which are the sources for the Glasma field.  In fact,
infinitesimally after the collision, the charge distribution on these
sheets is modified by the addition of sources of color electric and color
magnetic charge.  (We will discuss more precisely what is the physical
approximation which leads to this infinitesimal change, and what is
realistic physical time interval involved.)

The physical situation immediately after the collision bears close analogy 
to string models of high energy collisions~\cite{Andersson:1983ia,%
Ehtamo:1983hu,Biro:1984cf,Gatoff:1987uf}.
In  models such as that 
advocated by the Lund group, there is a longitudinal electric but no 
longitudinal magnetic field.
It is the decay of these fields which is the essential 
dynamics of the Glasma, and which ultimately produces the Quark Gluon 
Plasma.  For the Glasma, these fields can decay due to both classical 
rearrangement of the field, or from quantum pair creation.  The classical 
rearrangement of the field into radiation of gluons with $p_T \sim \qs$
 is a highly coherent process and happens on the 
time scales typical of the inverse separation of the color charges. 
The quantum process should be a factor of $\as$ weaker  and naively 
appears to a be a small correction to the classical process.
We shall argue that these processes are related.  We will discuss the recent 
proposal by Kharzeev, Levin and Tuchin that such fields are precisely the type 
which yield exponential distributions in transverse 
mass \cite{Kharzeev:2005iz,Kharzeev:2006zm}.

Finally, a longitudinal electric and magnetic field has a non-zero
topological charge density $F\widetilde{F}$.  We show how this results in a non-zero
Chern-Simons charge density.  Such a density can seed interesting
non-perturbative phenomena associated with chiral symmetry breaking. We
also argue that the decay of these fields bears a remarkable resemblance
to the picture advocated by Kharzeev, Kovchegov and Levin~\cite{Kharzeev:2001vs}
and by Janik, Shuryak and Zahed \cite{Janik:2002nk,Shuryak:2002qz},
which interprets high energy nuclear collisions as the
decay of instantons (see also 
Refs.~\cite{Schrempp:2005vc,Schrempp:2004vj,Schrempp:2002kd}
for a discussion of instantons in deep inelastic scattering).

\section{The Classical Equations}

For simplicity, we consider collisions at very high energy where the 
rapidity dependence of the distribution of produced particles can be 
ignored.  We assume the density of produced particles is so large that the 
typical transverse separation of the produced gluons is very small 
compared to a Fermi.  The typical QCD interaction strength is therefore
$\as \ll 1$.

We separate the degrees of freedom of a fast moving nucleus into large $x$ 
and small $x$ degrees of freedom.  This separation is arbitrary, but leads 
to an effective Lagrangian for the small $x$ degrees of freedom.  The 
parameters of this Lagrangian are subject to renormalization group 
evolution which ultimately resolves the ambiguity of the arbitrary scale 
separation, and determines the parameters of the effective Lagrangian.
The fast degrees of freedom are Lorentz contracted and on a sheet, and 
produce fields which ultimately describe the low $x$ degrees of freedom.

The fields of the single nucleus are the Li\'enard-Wiechert potentials
associated with the color charge distribution.  They exist also only in
the sheet and for each source of charge are mutually orthogonal electric
and magnetic fields which are also perpendicular to the beam direction. It
would appear that there is a paradox since the quanta associated with
these fields are at small $x$, although the spatial extent of the fields is
in the region of the valence quanta with a longitudinal size which is very
small.  By the uncertainty principle, it would seem that the gluons
associated with this field are at large $x$, not small. This paradox is
resolved by the fact that the small $x$ quanta are those of the
Fourier transform of the vector potential (in lightcone gauge). 
Although the color electric
and color magnetic field exist only within the sheet, the vector potential
has a discontinuity at the sheet but exists in the region outside the
sheet.  This is where the small $x$ gluons are located, so that the large
longitudinal spatial extent indeed corresponds to small $x$.

\begin{figure}[htbp]
\begin{center}
\includegraphics[width=0.70\textwidth]{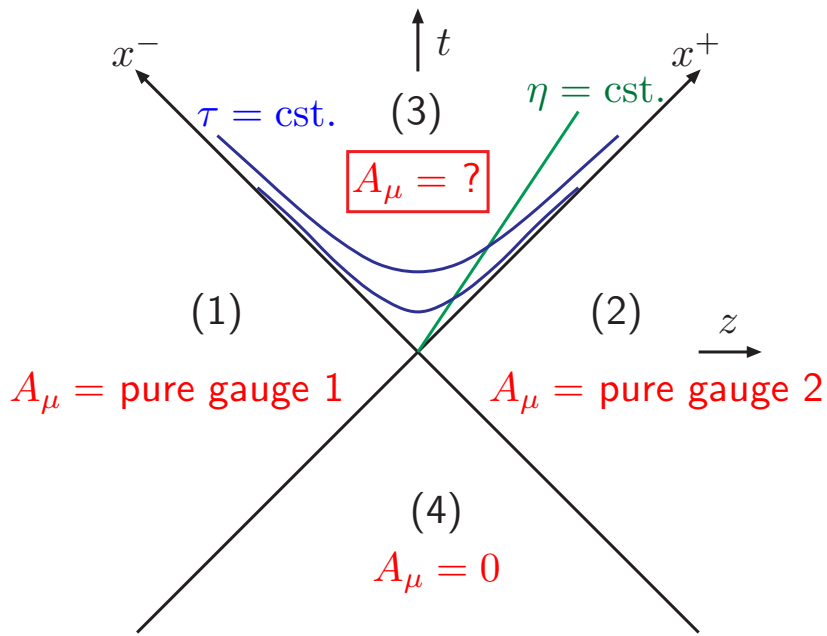}
\end{center}
\caption{Color fields in spacetime. In regions (1) and (2), where only one
of the nuclei has passed by, the field is the pure gauge field of this one
nucleus. In region (3) the field is known numerically.}
\label{fig:spacet}
\end{figure}

The sheet of color charge is approximated as infinitesimal, but its 
physical size can be estimated by renormalization group arguments.  
Let us concentrate on the nucleus moving along the $x^-=0$-lightcone.
There is a cutoff $\Lambda^+$ separating the hard and soft degrees
of freedom.
The degrees of freedom with $p^+ \gtrsim \Lambda^+$ are described
as classical color charges and the ones with $k^+ \lesssim \Lambda^+$ 
as classical fields, characterized by the saturation scale $\qs$.
The renormalization group equation for this system describes how
the distribution of color charges evolves when the cutoff is changed.
When $\Lambda^+$ changes by a factor $e^{\Delta y}$, 
the correction to the saturation
scale is of order $\as \Delta y$ and becomes of order one only when 
$\Delta y \sim 1/\as$. When studying the soft degrees of freedom around midrapidity
$y \sim \eta \sim 0$ we can therefore assume that they are separated
from the sources with $p^+ \gtrsim \Lambda^+$ by a rapidity difference 
$\Delta y \sim 1/\as$. The characteristic momentum and time scales for these soft modes
at $y \sim 0$ are $k^+ \sim \qs$ and $\Delta \tau \sim 1/\qs$. These
soft classical fields see hard sources as localized 
in an interval $\Delta z \sim \Delta x^- \sim 1/p^+ \lesssim e^{-1/\as}/\qs$
which, being
close to the $x^-=0$-lightcone (see Fig.~\ref{fig:spacet}), 
corresponds to a time interval $\Delta \tau \sim e^{-1/\as}/\qs$.
The smallness, in the weak coupling limit, of 
this scale $e^{-1/\as}/\qs$ compared to the scale $1/\qs$ of the soft fields 
is the physical picture approximated by the infinitesimally thin sheet.

For the collision of two nuclei, we imagine working in the center of mass 
frame.  Here the two nuclei can be thought of as sheets of colored glass.
The fields associated with the collision are illustrated in 
Fig.\ref{fig:spacet}.
We choose the fields to vanish in the backward light cone.  On the side 
light cones, we choose the fields to be two dimensional pure gauge 
transforms of vacuum,
\be
	A^i_{(1,2)} = \frac{1}{ig} U_{(1,2)} \nabla^i U^\dagger_{(1,2)} 
\ee
The derivatives are two dimensional and transverse in the sheets.
The discontinuity of the field at $x^{\pm} = 0$ in the backward light 
cone is determined by the source density on the two sheets,
\be \label{eq:src}
	J^\mu = \delta^{\mu +} \delta(x^-)\rho_1(\xt)
 +\delta^{\mu-} \delta(x^+) \rho_2(\xt)
\ee
In practice, to relate the vector potential to the charge density, one has 
to spread the charge density out in the longitudinal space, but this does 
not affect the determination of the solution of the classical field in 
terms of the vector potentials $A^i_{(1,2)}$.

In Fig. 2, the classical color and magnetic fields are shown on the two
nuclei before the collision.  They are frozen in time and the color 
electric and color magnetic fields are orthogonal to the beam direction 
and to one another.
\begin{figure}[htbp]
\begin{center}
\vspace{2cm}
\includegraphics[width=0.70\textwidth]{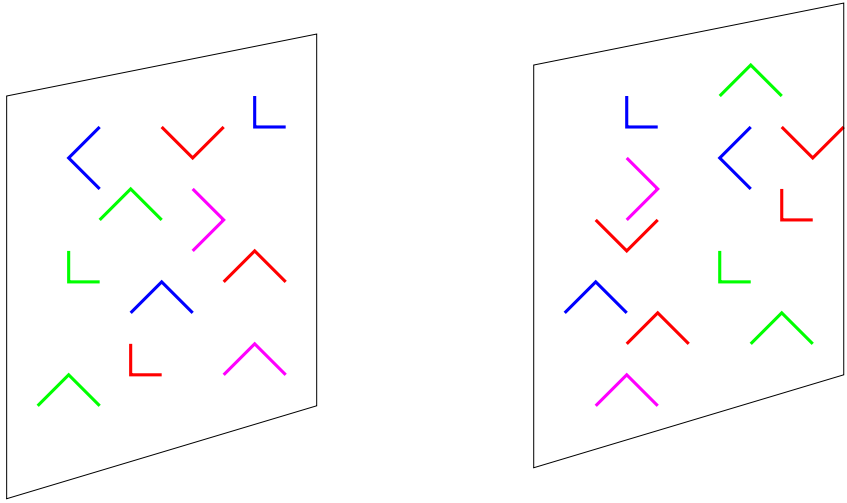}
\vspace{2cm}
\end{center}
\caption{The fields on the two nuclei prior to the collision.}
\label{fig:sheetonsheet}
\end{figure}

In the forward light cone, there is no solution which is pure gauge.  
There is therefore particle production and evolution in the forward 
lightcone.  This matter is the Glasma, until it eventually evolves into a 
thermalized Quark Gluon Plasma.  Because the sources, \eq\ref{eq:src}, are restricted to the light cone there is a boost invariant solution 
of the field equations in the forward light cone. It is convenient
to work in the $\tau,\eta$ coordinate system and in
radial gauge,
\be
	A_\tau = \frac{1}{\tau}\left(x^+A^- + x^-A^+ \right) = 0.
\ee
The field can then be written in terms of\footnote{The Hamiltonian equations 
of motion for the numerical calculation are often, e.g.
\cite{Krasnitz:1998ns,Lappi:2003bi} written in terms of 
$\phi \equiv A_\eta = -\tau^2 A^\eta$.} 
$\alpha \equiv A^\eta$ 
and the transverse components as
\be
	A^\pm & = &  \pm x^\pm \alpha(\tau, x_T) \nonumber \\
        A^i & = & \alpha_3^i(\tau,x_T).
\ee
The equations of motion for these fields were written explicitly in 
\cite{Kovner:1995ja,Kovner:1995ts}.
As second order partial differential equations they
require initial conditions for the fields and the first time derivatives.
These boundary conditions at $\tau = 0$ are determined by 
requiring that the Yang Mills equations be solved across the forward 
lightcone, and are
\be 
       \alpha_3^i \mid_{\tau = 0} &  = & \alpha^i_1 + \alpha^i_2
\nonumber \\
        \alpha \mid_{\tau = 0} & = & \frac{ig}{2} 
        			\left[\alpha^i_1,\alpha^i_2 \right]
\nonumber \\
		\dtau \alpha\mid_{\tau = 0} & = & \dtau \alpha_3^i\mid_{\tau = 0} =0.
\ee

\begin{figure}[htbp]
\begin{center}
\includegraphics[width=0.8\textwidth]{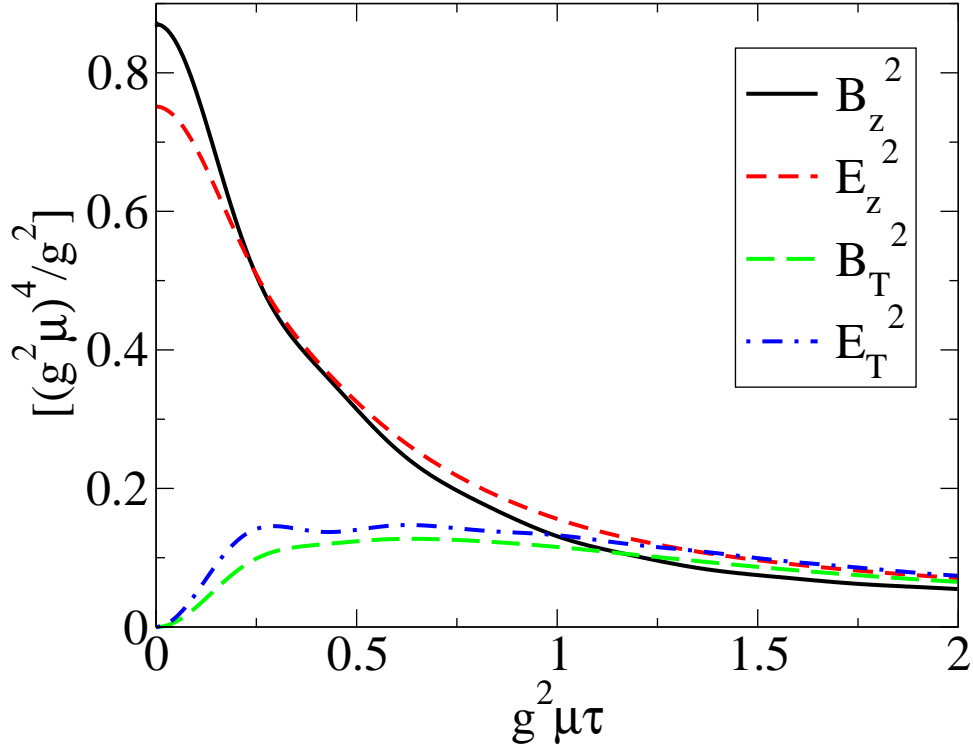}
\end{center}
\caption{Components of the gauge field, computed numerically on a 
$512^2$-lattice with $g^2\mu\ra = 67.7$.}
\label{fig:fields}
\end{figure}

If one then solves the Yang-Mills equation near the light cone, one
finds that the transverse color electric and color magnetic fields vanish 
as $\tau \rightarrow 0$, but that the longitudinal electric and magnetic 
fields\footnote{In the sense of the usual
$t,z$-coordinates, i.e. $E^z \equiv F^{tz}$ and 
$B^z \equiv F^{xy} = \half \epsilon^{ij}F^{ij}$.} are 
non-vanishing~\cite{Fries:2005yc,friesinprep},
\be\label{eq:longit}
	E^z & = & ig[\alpha^i_1,\alpha^i_2] \nonumber \\
    B^z & = & ig \epsilon^{ij} [\alpha^i_1,\alpha^j_2]. \\
\ee
A plot of the transverse and longitudinal color fields is shown in 
Fig.\ref{fig:fields}, 
based on a numerical solution of these equations.

A picture of the fields after the collision is shown in Fig. 4,
at a time infinitesimal after the collision.
In addition to the fields in the sheet, there are longitudinal electric 
and magnetic fields, which are random in color with a correlation length 
scale in the transverse plane of order $\mathcal{O}(1/\qs)$.
There are new sources of color electric and magnetic charge on the sheets on
which the original fields resided.

\begin{figure}[htbp]
\begin{center}
\includegraphics[width=0.8\textwidth]{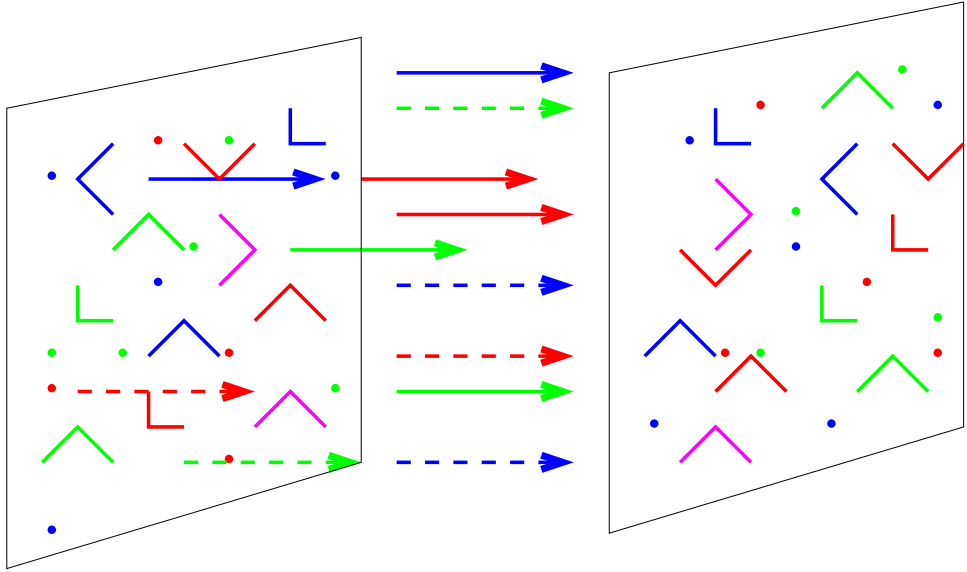}  
\end{center}
\caption{The color electric and magnetic fields after the collision.
In addition to the transverse fields on the sheets there is a longitudinal
field between them, originating from electric and magnetic 
charges on the sheets.}
\label{fig:sheetonsheet2}
\end{figure}

The presence of the longitudinal fields is at first sight surprising.  How 
did the longitudinal 
electric and magnetic field arise?  
Let us define the chromoelectric and chromomagnetic charge densities by
\be
	\nabla \cdot \vec{E} &  =  & \rho_{\mathrm{electric}} \nonumber \\
        \nabla \cdot \vec{B} & = & \rho_{\mathrm{magnetic}}.
\ee
Because the fields satisfy the nonabelian Gauss
law and the Bianchi identity these charge densities have a contribution
from the gauge field itself:
\be
	\rho_{\mathrm{electric}} &  = &  ig [A^i, E^i] \nonumber \\
        \rho_{\mathrm{magnetic}} & = & ig[A^i, B^i].
\ee

The source terms turn on instantaneously in the forward light cone.  This
arises because there are transverse electric and magnetic fields in the
sheets, which in the forward light cone are multiplied by a non-vanishing
vector potential from the opposing sheet.  The induced charge density on
one sheet is the negative of that on the other, as it must be to generate 
the longitudinal magnetic and electric fields and be consistent with 
Gauss's law.
To be explicit, the nucleus on the $x^-=0$ light cone has a gauge
potential $A^i = \theta(x^-)\alpha^i_1(\xt)$ and the other nucleus
$A^i = \theta(x^+)\alpha^i_2(\xt)$. The cross terms of these two give rise to 
the opposite charges on the light cones, electric
\be
\rho_{\mathrm{electric}}=
\frac{ig}{\sqrt{2}}\theta(x^-)\delta(x^+)[\alpha^i_1,\alpha^i_2] +
\frac{ig}{\sqrt{2}}\theta(x^+)\delta(x^-)[\alpha^i_2,\alpha^i_1] 
\ee
and magnetic
\be
\rho_{\mathrm{magnetic}}=
\frac{ig}{\sqrt{2}}\theta(x^-)\delta(x^+)\epsilon^{ij}[\alpha^i_1,\alpha^j_2] +
\frac{ig}{\sqrt{2}}\theta(x^+)\delta(x^-)\epsilon^{ij}[\alpha^j_1,\alpha^i_2],
\ee
which exactly correspond to the initial longitudinal fields, \eq\nr{eq:longit}.

After the collisions, sources of magnetic and electric charge have
appeared on the sheets.  The valence structure of the nuclei have
changed, and there are new sources for the fields.  This description is
very similar in spirit to the original suggestions for a string based
description of heavy ion collisions. These old pictures had only sources
of electric charge~\cite{Andersson:1983ia,%
Ehtamo:1983hu,Biro:1984cf,Gatoff:1987uf}.
The appearance of a longitudinal magnetic field is an
entirely new aspect.

In the Lund Monte Carlo model~\cite{Andersson:1983ia}, 
there are longitudinal electric fields 
induced by a collision, which subsequently decay by quantum pair 
production.  The situation here is similar except for the longitudinal 
magnetic field and the different length scale in the transverse direction,
$1/\qs$ in stead of $1/\lqcd$. 
The other essential difference is that the fields can 
decay away by classical evolution of the charged Yang-Mills field.  There 
is not a necessity for pair production from quantum effects to quench the 
fields.  We will expand on this later, since, the quantum and classical 
quenching processes are related in a non-trivial way.

It is interesting to note the structure of the energy momentum tensor 
$T_{\mu \nu} = \frac{1}{4}g_{\mu \nu}F^{\alpha \beta}F_{\alpha\beta}
- F_\mu{}^\alpha F_{\nu \alpha} $ for this initial condition. It is diagonal
and, as always in gauge theory, traceless: 
$T_{\mu \nu} = \half (E_z^2 + B_z^2) \times \mathrm{diag}(1,-1,-1,1)$. This can be
compared to the standard form for a system with an anisotropy in the $z$-direction:
$T_{\mu \nu}= \mathrm{diag}(\epsilon,-p_\perp,-p_\perp,-p_L),$
where $\epsilon$ is the energy density and
$p_\perp$ and $p_L$ are the transverse and longitudinal pressures. We see
that the initial field configuration has \emph{negative} ``longitudinal pressure''.
The configuration that is the starting point for studies of isotropization
by plasma instabilities, where the diagonal elements of $T_{\mu\nu}$ at
$\eta=0$ are $(\epsilon,-\epsilon/2,-\epsilon/2,0)$, is only reached at times
$\tau \gtrsim 1/\qs$ when the classical fields start to behave linearly due
to the expansion of the system.

Finally, the non-zero longitudinal electric and magnetic fields implies 
that the initial conditions have a large density of topological charge 
$F^{\mu\nu} \widetilde{F}_{\mu\nu}$.  Since
\be
  \partial^\mu K_{\mu} = F^{\mu\nu}\widetilde{F}_{\mu\nu}
\ee
there is also a nonzero Chern-Simons current.  We will investigate the 
topology of these configurations in the next section.  The physical 
picture we have of the initial conditions looks quite similar to that 
proposed by Kharzeev, Kovchegov and Levin~\cite{Kharzeev:2001vs}
and by Janik, Shuryak and Zahed~\cite{Janik:2002nk,Shuryak:2002qz}, 
who claimed  that high energy collisions might be described by the decay of 
instanton like configurations, the sphalerons.  The decay of these 
longitudinal electric and magnetic fields is in fact the decay of 
topological charge.

To thermalize the system, one must first produce quanta associated with 
the gluon field.  In order to be described as quanta, the classical field 
associated with these quanta must be weak, $A \ll 1/g$.  For a mode with 
momentum $p_T$, this occurs in a time of order $\tau \sim 1/p_T$.  For 
a time $1/p_T \ll \tau \ll 1/(\alpha_S p_T)$,
the field has strength $1 \ll A \ll 1/g$, 
and can be treated either classically, or as an ensemble of 
particles.  This follows because, up to oscillations, the field behaves as
$1/\sqrt{\tau}$ at late times.  For $p_T \gg 1/(\alpha_S \tau)$, the field 
is essentially quantum, as the classical field is small compared to the field 
induced by quantum fluctuations (see~\cite{Baier:2000sb} for
an argument approaching this same limit from the opposite direction, i.e.
kinetic theory, and~\cite{Mueller:2002gd,Jeon:2004dh} for a more
formal argument).

Roughly speaking, the emission of particles occurs during the time 
interval $1/\qs \ll \tau \ll 1/(\as \qs)$.  Since $\as$ varies 
logarithmically with $\qs$, the time scale is crudely $1/\qs$.  
Nevertheless, this treatment suggests there is an intermediate time scale 
where either the classical and transport solutions to the evolution of a 
coupled 
system of fields and particles may be a good approximation.  Above the 
high momentum end of 
this matching interval, one should use a hard particle treatment.  Below 
the lower scale of the overlap region, one should use a classical field 
treatment.  For a complete treatment, one should have a floating scale 
which depends upon time.  Modes above this floating scale are hard 
particles, and those below are soft.  

The picture is therefore of classical fields evaporating into gluons,
and a system where the hard gluons are interacting with a classical 
field.
The interaction of the hard gluons among themselves is probably not so 
important during the time interval $1/\qs \ll \tau \ll 1/(\as \qs)$ 
since the characteristic time scale for hard particles to 
thermalize by interactions among themselves is of order $\tau \sim 
1/(\as^2 \qs)$, since these interactions involve scattering cross 
sections.  The interaction of the classical field with itself is 
important, and the characteristic time scale for interaction of the hard 
fields with the classical fields is also important, since the effects of 
coupling constant cancel out in the interactions of the hard field with 
the coherent fields strength $A \sim 1/g$.  

It is this system which we refer to as the Glasma.    

An issue of current interest is whether or not the boost invariant 
solution described in the previous section is stable with respect to space-time 
rapidity dependent fluctuations.  There is reasonably strong evidence that 
the coupled particle--field Glasma system has such instabilities.  The 
issue of the purely classical evolution also has claims that this may be 
the case~\cite{Romatschke:2005pm},
but they happen over extremely long time scale, and one may be 
worried about whether such instabilities reflect the continuum limit.

As we will discuss in Sec.~\ref{sec:topo}, the topological properties of the gauge fields depend crucially on the dimensionality of the system. One would therefore expect that relaxing the strict boost invariance
of the field configurations will significantly
change the dynamics of the Chern-Simons charge.
Perhaps the instability of the boost invariant Glasma and the production 
of net topological charge might be related.  

If the purely classical system is unstable, then one may legitimately 
worry about the validity of the entire Glasma description. The problem is 
that if the instability is like that of classical turbulence, then 
different initial conditions which are separated from one another 
infinitesimally, will at late times evolve to solutions far apart 
with $\Delta S \sim S$ .  Such a
difference in initial conditions might be generated by quantum noise.
Their contribution to the path integral will differ by a factor of
\be
	e^{i\Delta S}
\ee
which is oscillatory.  Therefore one would not expect convergence of the 
glassy sum over initial conditions.

\begin{figure}[htbp]
\begin{center}
\hfill
\includegraphics[width=0.35\textwidth]{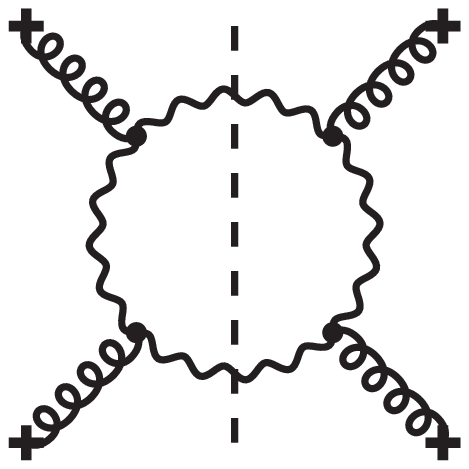}  
\hfill
\includegraphics[width=0.35\textwidth]{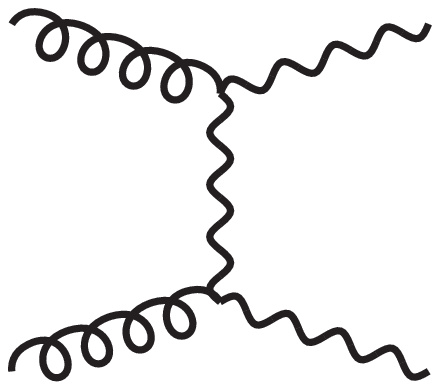}  
\hfill
\rule{0pt}{1pt}
\end{center}
\caption{Left: A lowest order (both in the classical field and 
in the quantum corrections) diagram that
contributes to the imaginary part of the effective action.
The winding lines represent the classical field, $\mathcal{O}(1/g)$
and the wavy lines the quantum fluctuation, $\mathcal{O}(1 )$.
Right: The corresponding pair production diagram. }
\label{fig:effact}
\label{fig:tchan}
\end{figure}

There can also be instabilities with $\Delta S \sim \as S \ll S$.
In addition to the interpretation of the momentum modes of
the classical field as gluonic quanta there is 
quantum radiation.  This may be generated by considering the small 
fluctuations in the classical time dependent background field. The 
effective action is then obtained by integrating over these fluctuations.
 At fourth order in the expansion of the effective action in powers of the 
external field, there is an imaginary part generated, coming from 
diagrams such as the one in Fig.~\ref{fig:effact}.
This imaginary part is precisely pair production.
Pair production does not correspond to an instability of the purely classical 
equations, but to an instability of the classical 
background field in a quantum
mechanical treatment. It remains an effect suppressed by $\as$ with
respect to the classical background and is therefore not a problem for the 
validity of the glassy description. 
A consistent computation of this pair production must also include the
renormalization group evolution of the classical fields, which contributes 
at the same order in $\as$~\cite{Gelis:2006yv}.
Kharzeev, Levin and Tuchin~\cite{Kharzeev:2005iz,Kharzeev:2006zm}
have argued that for fields  of the type we have shown exist in the 
collision of two sheets of colored glass, specifically a pulsed 
longitudinal electric field, the distribution of produced particles is an 
exponential in transverse mass, contributing to the early thermalization
of the system.

The contribution to the gluon multiplicity from pair production
is related to the diagram shown in 
Fig.~\ref{fig:tchan}, where one integrates over one of the legs of the quantum 
fluctuation field. 
In this integration the integral over $\Delta y$, the rapidity gap between
the produced gluons, diverges (see e.g. Appendix B of \cite{Leonidov:1999nc}).
The region $\Delta y \gtrsim 1/\as$ corresponds to renormalization group
evolution of the sources and the genuine quantum correction to
gluon production only comes from the region $\Delta y \ll 1/\as$.
To compute the first quantum correction to gluon production
one must introduce a cutoff $Y \sim 1/\as$ in  the integration over $\Delta y$. 
One then integrates over one of the legs in Fig.~\ref{fig:tchan} up to this 
cutoff. The physical result is independent of $Y$, because the dependence
of the gluon multiplicity on $Y$ must be cancelled by the renormalization 
group evolution of the sources as a function of $Y$. Thus pair production
remains suppressed by a power of $\as$ compared to the background field.
This can be contrasted with the calculation of quark pair production
\cite{Gelis:2003vh,Fujii:2005vj,Gelis:2004jp,Gelis:2005pb}, where the
integral over $\Delta y$ is convergent and the quark multiplicity
is directly suppressed by $\as$ compared to the classical field. Quark pairs, 
unlike gluons, do not contribute to the renormalization group evolution of
the source to this order.

\section{Topological Charge} \label{sec:topo}

The topological charge density of a gauge field configuration is defined as
\be \label{eq:fftilde}
\frac{\ud^4 \nu}{\ud x^4} 
= \frac{g^2}{32 \pi^2} \Tr \widetilde{F}^{\mu\nu}F_{\mu\nu} 
=  \frac{g^2}{8 \pi^2} \Tr \vec{E} \cdot \vec{B},
\ee
where $\widetilde{F}^{\mu\nu} = \half \epsilon^{\mu\nu\rho\sigma} F_{\rho\sigma}$.
The charge density is the four-divergence of the topological current,
$\widetilde{F}_a^{\mu\nu}F^a_{\mu\nu} = \partial_\mu K^{\mu}$, where 
\be
K^\mu = \epsilon^{\mu \nu \rho \sigma} A^a_{\nu}
\left( F^a_{\rho\sigma} + \frac{g}{3} f^{abc}A^b_\rho A^c_\sigma \right).
\ee
Integrating the 0-component of the current over space one can define a topological charge
$\nu = \frac{g^2}{64\pi^2} \int \ud^3 \mathbf{x} K^0$.
 The change in the topological charge between times,
\be
\nu(t\to\infty)-\nu(t\to -\infty) = \int \ud t \partial_t \nu 
=  \frac{g^2}{32 \pi^2} \int \ud^4 x \Tr \widetilde{F}^{\mu\nu}F_{\mu\nu},
\ee
is a gauge invariant quantity, but the topological current and charge 
themselves are gauge dependent. It is therefore essential in our case
that one cannot transform away the pure gauge contribution at large $\tau$,
because such a gauge transformation would introduce in the backward lightcone
a nonzero gauge potential $A_i$.

Working in the boost invariant case it is more natural to look at the
Chern-Simons charge per unit rapidity on a constant proper time surface
\be\label{eq:dnudeta}
\frac{\ud \nu}{\ud \eta} = \frac{g^2}{64\pi^2} \int \ud^2 \xt \tau K_\tau
= \frac{g^2}{64\pi^2} \int \ud^2 \xt 
\epsilon^{ij}\left[ A^a_\eta \left(F^a_{ij} + g f^{abc}A^b_i A^c_j\right) 
+ 2 A^a_i F^a_{j\eta}
\right]
\ee
As was pointed out in \cite{Kharzeev:2001ev} this expression can be 
written in a much simpler form when the field configurations are
boost invariant. Performing a partial integration one can rearrance 
Eq.~\nr{eq:dnudeta} as
\be \label{eq:kkvncs}
\frac{\ud \nu}{\ud \eta} = \frac{g^2}{32\pi^2} \int \ud^2 \xt 
\epsilon^{ij} A^a_\eta F^a_{ij} = 
\frac{g^2}{16 \pi^2} \int \ud^2 \xt  A^a_\eta B^{az}.
\ee
The surface term from the partial integration is
\be \label{eq:surfncs}
\frac{g^2}{32 \pi^2} \int \ud^2 \xt \epsilon^{ij} \partial_j 
\left( A^a_i A^a_\eta \right)
= \frac{g^2}{32 \pi^2} \oint \ud x^i  A^a_i A^a_\eta.
\ee

There should be a large topological charge density 
associated with the Glasma.  This is because the Glasma fields at 
$\tau \to 0^+$ have longitudinal $\vec{E}$ and $\vec{B}$ 
and give nonvanishing $\vec{E} \cdot \vec{B}$. 
Of course, the typical correlation length in these fields
in transverse size is the saturation momentum, and for reasons described 
below, we expect zero total Chern-Simons charge from these fields.  It is 
interesting that this resembles the situation envisioned by
Kharzeev, Kovchegov and Levin~\cite{Kharzeev:2001vs} and Janik, Shuryak and 
Zahed~\cite{Janik:2002nk,Shuryak:2002qz} 
in that the production of particles can be thought 
of as due to the decay of Chern-Simons charge.  Presumably, this is 
somewhat like the situation at finite temperature where one has a density 
of sphalerons associated with the decay of topological charge.  It is 
amusing here that the topological charge density arises already at
the classical level and is $\mathcal{O}(1)$, not $\mathcal{O}(\as)$
as it would be from quantum fluctuations.

For early times the electric and magnetic fields are both longitudinal, and
there is a nonzero topological charge density. Note that because 
$\ud /\ud \eta = \tau \ud / \ud z$ the charge density per unit rapidity 
is zero, but the density per unit $z$ nonzero. 
The expectation value of the Chern-Simons charge density when averaged over
the source color charge densities is zero, but the magnitude
of the fluctuations, 
$\sqrt{\left\langle  \left( \ud\nu/\ud \eta \right)^2 \right\rangle}$,
can be estimated as follows. The system consists of approximately
$N=\pi\ra^2\qs^2$ uncorrelated domains, each having charge $\sim 1$.
The sum of these independent charges will fluctuate around zero with magnitude
$\sqrt{ \left\langle  \left( \ud\nu/\ud \eta \right)^2 \right\rangle} 
\sim \sqrt{N} \sim \qs \ra$. This estimate is confirmed by the numerical 
calculation \cite{Kharzeev:2001ev}.

For late times $\tau \to \infty$ the field gets weak due to the expansion of
the system and the gauge field can be written as a gauge transformation of 
a solution to the linearized equations of motion~\cite{Kovner:1995ts}. 
The general such asymptotic form of the solution is:
\be \label{eq:asymp}
A_i &=& V(\xt) \left(\epsilon_i(\xt,\tau) - \frac{i}{g} 
\partial_i \right)V^\dag(\xt) \nonumber \\
A_\eta &=& V(\xt)\epsilon(\xt,\tau)V^\dag(\xt)
\ee
with
\be
\epsilon_i(\xt,\tau) &=& \int \frac{\ud^2\kt}{(2 \pi)^2} e^{i \kt \cdot \xt}
\left[ J_0(|\kt|\tau) h^J_i(\kt) + N_0(\kt\tau) h^N_i(\kt) \right]
\nonumber \\
\epsilon(\xt,\tau) &=&  \int \frac{\ud^2\kt}{(2 \pi)^2} e^{i \kt \cdot \xt}
\tau \left[ J_1(|\kt|\tau) h^J(\kt) + N_1(\kt\tau)h^N(\kt) \right].
\ee
This decomposition between the field and the pure gauge contribution
is not unique, but this ambiguity can be removed by requiring that
$\epsilon_i$ satisfies the transversal Coulomb gauge condition
$\partial_i \epsilon_i=0$.
The mode functions $h^J_i(\kt), h^N_i(\kt),h^J(\kt),h^N(\kt)$ are determined
by the nonlinear dynamics of the system at early times.

In the lowest order  perturbative solution \cite{Kovner:1995ts} the mode 
functions are determined only by the initial conditions. Because the Neumann
functions $N_n$ diverge at the origin, $h^N_i(\kt)$ and $h^N(\kt)$ vanish
at in the lowest order. Thus the Chern-Simons charge, \eq\ref{eq:kkvncs}, for 
each transverse momentum mode behaves as $J_0(|\kt|\tau)J_1(|\kt|\tau)$,
giving a zero contribution when averaging over a time period $\tau \gg 1/|\kt|$.
At higher orders in the source also the Neumann function solutions can be 
present and $h^N_i(\kt)$ and $h^N(\kt)$ can be finite. Because 
$J_0(|\kt|\tau)N_1(|\kt|\tau)$ and $N_0(|\kt|\tau)J_1(|\kt|\tau)$ oscillate
around finite values, these terms can give a finite radiative contribution
to the topological charge. This higher order radiative contribution is
the most likely interpretation for the finite topological charge seen in 
the numerical calculation \cite{Kharzeev:2001ev}.

When, as in Ref.~\cite{Kharzeev:2001ev}, the topological charge 
is computed on a finite lattice 
with periodic boundary conditions and thus the surface term \eq\nr{eq:surfncs}
vanishes exactly. In general it is possible to have a solution
where the combination of a covariantly constant longitudinal field
$A_\eta = V(\xt)\epsilon V^\dag(\xt)$ with 
$\dtau \epsilon = \partial_i \epsilon = 0$ and the pure gauge component 
$V(\xt)$ give a nonvanishing
topological charge. This contribution is not present in the perturbative
solution of Ref.~\cite{Kovner:1995ts}, because a constant
$A_\eta$ corresponds to $A^\eta$ that diverges at $\tau=0$ and is not
allowed by the initial condition. For an arbitrary color charge distribution
in the transverse plane the pure gauge fields of the individual nuclei, and
thus also the pure gauge fields in the asymptotical solution \eq\nr{eq:asymp},
vanish only logarithmically for large $\xt$. However the overall color
charge of the nucleus must be zero, and thus for distances larger
than $1/\lqcd$ outside the nucleus the color field must die off faster,
meaning that the boundary term will vanish.

\subsection{Homotopy Groups}

The $n$th homotopy group $\pi_n(G)$ of a topological space $G$, 
in our case the gauge or symmetry group, is
the group of mappings from the $n$-sphere $S^n$ to $G$. The usefulness of
homotopy groups in field theory often arises in the following way. 
Let us presume were are interested in a field theory with symmetry group $G$.
When one requires that field configurations in $R^n$ approach a constant at 
large distances, one can view space as the compact group $S^n$. The homotopy 
group $\pi_n(G)$ then tells us whether topologically inequivalent field 
configurations exist. One example is the group 
$\pi_2(\textrm{O(3)}) = \mathbb{Z}$, which tells us that in an O(3)-symmetric
nonlinear sigma model in 2 dimensions
one can construct a topological charge that takes integer
values \cite{Mottola:1988ff}. Another example is the Skyrme 
model~\cite{Skyrme:1961vq,Adkins:1983ya}, where the 
solitons of an SU(2) symmetric ``pion'' field in three dimensions
can be interpreted as having integer baryon number, because 
$\pi_3(\textrm{SU(2)})= \mathbb{Z}$. For this same reason the Chern-Simons
number of SU(3) in a gauge theory changes by an integer in a nonperturbatively
large gauge transformation. The second homotopy group of SU(3), however, is trivial:
$\pi_2(\textrm{SU(3)})=0$, meaning that for
boost invariant two dimensional field configurations one can not construct a 
topological charge taking integer values.

\section{Conclusions}

We have in this paper tried to elucidate some known but little appreciated
aspects of Glasma, the highly coherent matter making in transition from 
Color Glass Condensate to Quark Gluon Plasma. It is characterized, in
a manner reminiscent of the Lund model, by the decay of a longitudinal
electric and magnetic field into particles. This decay can happen both 
as classical radiation of the field and, quantum pair production from the 
classical background. We have discussed the relation of 
these two processes and pointed out that a full computation of the
pair production must be done consistently with
the renormalization group evolution of the sources.

As the magnetic and electric fields are initially parallel, they correspond
to a nonzero Chern-Simons charge density. The evolution of the Glasma can thus 
be described also as the decay of this Chern-Simons charge.

\section*{Acknowledgements}

One of the authors, L. McLerran wishes to thank the late Bo Andersson, 
who insisted that there should be longitudinal color electric fields
in hadronic collisions.  L. McLerran insisted at the time that QCD should 
have only transverse fields associated with the boosted charge 
distributions of the color degrees of freedom of a hadron. 

Bo was correct.

L. McLerran wishes to thank Rainer Fries, Joe Kapusta and Yang Li for
crucial observations about the existence of such longitudinal electric
fields at early times.  L. McLerran also acknowledges the gracious hospitality
of the University of Minnesota where these discussions took place.
T.L. is thankful to K. Kajantie for insisting that there is interesting physics
yet to be understood in the initial condition for the gauge fields. We both
wish to thank Dima Kharzeev and Raju Venugopalan for sharing their wisdom
about hadron collisions within the theory of the Color Glass Condensate
and Yoshitaka Hatta for many discussions on this paper.
This manuscript has been authorized under Contract No. DE-AC02-98CH10886 
with the U. S. Department of Energy.

\bibliographystyle{h-elsevier3}
\bibliography{spires}

\end{document}